\documentclass[astronomy,article,accept,pdftex,oneauthor]{Definitions/mdpi} 

\firstpage{1} 
\pubvolume{4}
\issuenum{2}
\articlenumber{6}
\pubyear{2025}
\copyrightyear{2025}
\externaleditor{Firstname Lastname}
\datereceived{28 January 2025} 
\daterevised{14 March 2025}
\dateaccepted{31 March 2025} 
\datepublished{ } 
\hreflink{https://doi.org/}
\usepackage{braket}

\Title{Distance to M87 as the Mode of the Modulus Distribution}

\TitleCitation{Distance to M87 as the Mode of the Modulus Distribution}

\Author{Mariusz Tarnopolski \orcidA{}}

\AuthorNames{Mariusz Tarnopolski}

\isAPAStyle{%
       \AuthorCitation{Tarnopolski, M.}
         }{%
        \isChicagoStyle{%
        \AuthorCitation{Tarnopolski, Mariusz}
        }{
        \AuthorCitation{Tarnopolski, M.}
        }
}

\address[1]{Institute of Astronomy, Faculty of Physics, Astronomy and Informatics, Nicolaus Copernicus University, \linebreak ul. Grudzi\k adzka 5, PL-87-100 Toru\'n, Poland; mariusz.tarnopolski@umk.pl}

\abstract{de Grijs and Bono (ApJS 2020, 246, 3) compiled
 a list of distances to M87 from the literature published in the last 100 years. They reported the arithmetic mean of the three most stable tracers (Cepheids, tip of the red giant branch, and surface brightness fluctuations). The arithmetic mean is one of the measures of central tendency of a distribution; others are the median and mode. The three do not align for asymmetric distributions, which is the case for the distance moduli $\mu_0$ to M87. I construct a kernel density distribution of the set of $\mu_0$ and estimate the recommended distance to M87 as its mode, obtaining $\mu_0 = \left(31.06~\pm~0.001\,\textrm{(statistical)}\,^{+0.04}_{-0.06}\,\textrm{(systematic)}\right)$~mag, corresponding to \linebreak $D=16.29^{+0.30}_{-0.45}$~Mpc, which yields uncertainties smaller than those associated with the mean and median.}

\keyword{
distance scale cosmology; distance and redshift galaxies; data analysis method; statistical methods}

\begin{document}

\section{Introduction}
\label{sect:1}

\citet{degrijs20} compiled a list of more than 200 distance moduli $\mu_0$ (absorption-corrected) 
to M87 from the literature published in the last 100 years. They categorized them into 15 groups based on the employed tracers and employed the five they established to be internally consistent: period--luminosity relation for Cepheids, planetary nebulae luminosity function (PNLF), surface brightness fluctuation (SBF), the tip of the red giant branch (TRGB) magnitude, and novae. Fundamentally, all of these techniques rely on calibrating a given type of astronomical sources or phenomena as standard candles owing to some of their properties being universal across cosmic time:
\begin{enumerate}
\item Cepheids are periodically varying stars whose period is tightly related to their intrinsic luminosity (Leavitt law \citep{leavitt}) and, hence, can be employed to infer the distances to their host galaxies.
\item  PNLF \citep{ciardullo98}---the LF of PN is very consistent across different galaxies, and the number and luminosity of PN can be used to infer the distance via a calibrated intrinsic luminosity of PN. The PNLF is independent of the galaxies' types and environments.
\item  SBF \citep{tonry88}---small fluctuations of surface brightness due to unresolved stars in a galaxy are used to infer its distance; especially applicable to nearby ellipticals, such as M87.
\item  The TRGB \citep{beaton18}
in a galaxy's color--magnitude diagram corresponds to a known, calibrated luminosity and as such can be used to infer the distance to the host galaxy.
\item  The peak brightness of novae \citep{valle95} has a consistent luminosity (similarly to supernovae) and, hence, can be calibrated to serve as a distance indicator.
\end{enumerate}

The adjusted distance values to M87 from the PNLF were eventually discarded as being essentially lower limits, and novae had significantly larger uncertainties. Using Cepheids, SBF, and TRGB, the arithmetic mean of $\mu_0$ was calculated as $(31.03\pm 0.14)$~mag, corresponding to $D=(16.07\pm 1.03)$~Mpc. Note that despite calculating the ``mean values and their 1$\sigma$ uncertainties'', the uncertainty was, in fact, taken as the standard deviation, not as the standard error of the mean, which should be divided by the square root of the sample size (i.e., by $\sqrt{24}$) and, hence, should read 0.03~mag or 0.21~Mpc.
\footnote{See the discussion in Section~\ref{sect:3.1} for an explanation for the factor of 24 instead of 25, as Section~\ref{sect:2.1} would suggest.}

The arithmetic mean is one possible measure of central tendency of a distribution. Another option is the median, which was employed recently by~\citet{rackers23}, who used the full sample of~\citet{degrijs20} to arrive at a value of the systematic error, and using the same sample of Cepheid, SBF, and TRGB moduli gave the median distance as $31.08^{+0.06}_{-0.08}$~mag, corresponding to $D=16.4^{+0.5}_{-0.6}$~Mpc.

\textls[15]{\citet{ramakrishnan23} reported the median of all 15 tracers to be} \linebreak $\mu_0 = (31.08\pm 0.09)$~mag, corresponding to $D=(16.44\pm 0.68)$~Mpc. They examined the Gaussianity of the residuals and concluded that whereas some tracers fulfill this condition (e.g., PNLF and SBF), the full set of moduli does not; therefore, the mean is unsuitable and the median ought to be employed as the distance to M87.

The third common measure of central tendency is the mode, i.e., the most probable value that manifests itself by a peak in the probability density function (PDF). I utilize the same sample of~\citet{degrijs20} to establish the distance to M87 as the mode of the distribution of $\mu_0$ to complement the previous studies with a measure of central tendency that has not yet been employed to such a task. Bootstrap \citep{efron1,efron2,efron3} is employed to estimate the statistical error. The final distance $D=16.29^{+0.30}_{-0.45}$~Mpc is consistent with the other two estimates (i.e, the mean and median), and including the systematic error from \citep{rackers23} leads to even smaller uncertainties than in the other studies. 

\section{Data and Methodology}

\subsection{Data}
\label{sect:2.1}

\begin{figure}[ht!]
\centering
\includegraphics[width=0.495\columnwidth]{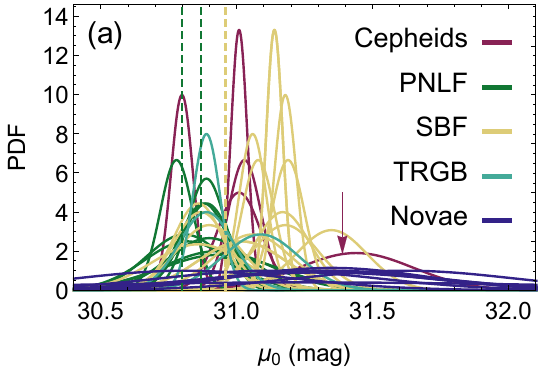}
\includegraphics[width=0.495\columnwidth]{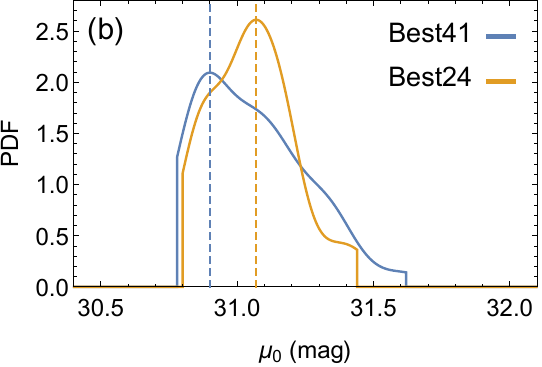}
\caption{(\textbf{a}) Distance moduli of the measurements from \citep{degrijs20} represented as Gaussian distributions with locations $\mu_0$ and their individual standard errors as standard deviations. Vertical dashed lines mark values with no errors reported. Arrow points to the~\citet{tammann2000} measurement (see Section~\ref{sect:3}). (\textbf{b}) Distributions of the investigated samples. Vertical dashed lines mark the respective modes; their difference equals 0.17~mag.}
\label{fig1}
\end{figure}

The distance moduli $\mu_0$ are taken from Table 1 of \citep{degrijs20}. They are grouped according to tracers: Cepheids, PNLF, SBF, TRGB, and novae, with sample sizes of $N=5, 11, 17, 3, 8$, respectively. They are represented in Figure~\ref{fig1}a as Gaussian distributions, $\mathcal{N}(\mu_0,\sigma)$, with individual values $\mu_0$ and their errors $\sigma$ as means and standard deviations. Combined, they form the Best44 sample. Two PNLF distances and one SBF distance are given with no errors by the originating authors. Excluding these gives the Best41 sample. Taking only Cepheids, SBF, and TRGB as the most reliable tracers from the Best44 sample forms the Best25 sample. Excluding the one SBF value with no error gives the Best24 sample.

\subsection{Methods}
\label{sect:2.2}

The final two samples from Section~\ref{sect:2.1} are displayed in Figure~\ref{fig1}b as kernel density estimates (KDEs) with a Gaussian kernel and bandwidth according to the~\citet{silverman86} rule. The KDEs are unimodal; hence, the mode unambiguously corresponds to the most probable value. I note in passing that the modes of the Best44 and Best41 samples are close to each other, as well as the modes of the Best25 and Best24 samples. This means that the $\mu_0$ values not equipped with uncertainties in the originating papers do not strongly affect the resulting distance distributions; nevertheless, the Best44 and Best25 samples are excluded from the following analyses due to the inability to account for the biases the missing uncertainties introduce. There is a discrepancy between the modes of the samples consisting of all five tracers and the most stable three, amounting to 0.17~mag. PNLF values being essentially lower limits (PN detections are biased toward foreground objects \citep{degrijs20}) offsets the modulus distribution toward lower modes; indeed, the Best24 sample has its mode higher than the Best41 sample. 

To constrain the positions and uncertainties of the modes, bootstrap resampling \citep{efron1,efron2,efron3} is performed as follows:
\begin{enumerate}
\item From the Gaussian representation $\mathcal{N}(\mu_0,\sigma)$ of each measurement in Best41 and Best24 samples a randomized value is drawn, thus creating a random realization of the Best41 and Best24 samples;
\item KDEs of the randomized Best41 and Best24 samples are constructed, and their modes $m$ are found numerically;
\item The above procedure is repeated $n=10^4$ times to form distributions of modes $\{m_i\}_{i=1}^n$; their means $\braket{m}$, standard deviations $\sigma_m$, and standard errors of the mean $\sigma_m/\sqrt{n}$ \linebreak are computed;
\item The above procedure is repeated independently 10 times to verify that $\braket{m}$ and $\sigma_m$ always converge to approximately the same values.
\end{enumerate}

\section{Results}
\label{sect:3}

\subsection{Discussion}
\label{sect:3.1}

The outcomes of the bootstrap from Section~\ref{sect:2.2} are gathered in Table~\ref{tab1}. Mean mode $\braket{m}$ for the Best41 sample is slightly smaller than for the Best24 sample. The standard deviations and standard errors of the means are comparable between samples. It should be emphasized that the standard errors of the sample means are small because they scale with $1/\sqrt{n}$. The standard error of the mean is not a measure of dispersion of a sample, which, even if large, does not reflect the accuracy of determining the {\it value of the mean} whose constraint becomes tighter for larger samples. This was correctly computed by~\citet{rackers23}; cf. their Table~1.

\begin{table}[H]

\caption{Mean $\braket{m}$ values of the bootstrap resampled distributions of mode $m$, their standard deviation $\sigma_m$, standard error $\sigma_m/\sqrt{n}$ of the means, distances $D$, statistical errors and final errors $\Delta D^-$ and $\Delta D^+$ obtained as statistical errors combined in quadrature with systematic errors from \citep{rackers23}. Quantities are rounded to 0.01, except for $\sigma_m/\sqrt{n}$, which is rounded to 0.001.}
\label{tab1}
\begin{tabularx}{\textwidth}{LCCCCCCC}
\toprule

	\multirow{2}{*}{\textbf{Sample}}	 & \boldmath{$\braket{m}$} & \boldmath{$\sigma_m$} & \boldmath{$\sigma_m/\sqrt{n}$} & \boldmath{$D$} & \boldmath{$\Delta D^\textrm{stat}$} & \boldmath{$\Delta D^-$} & \boldmath{$\Delta D^+$} \\
            & \textbf{(mag)} & \textbf{(mag)} & \textbf{(mag)} & \textbf{(Mpc)} & \textbf{(Mpc)} & \textbf{(Mpc)} & \textbf{(Mpc)} \\
		\midrule
		Best41 & 31.00 & 0.07 & 0.001 & 15.85 & 0.01 & 0.44 & 0.29 \\
            Best24 & 31.06 & 0.06 & 0.001 & 16.29 & 0.01 & 0.45 & 0.30 \\
		\bottomrule
	\end{tabularx}
\end{table}

The distance $D$ is calculated as follows: 
\begin{equation}
D=10^{1+\mu_0/5}
\label{eq1}
\end{equation}
with $\mu_0$ taken as the mean of the mode distribution from bootstrapping $\braket{m}$ (Section~\ref{sect:2.2}). The error $\Delta D$ is obtained via error propagation as
\begin{equation}
\Delta D = \frac{D}{5}\ln(10)\Delta\mu_0.
\label{eq2}
\end{equation}
These statistical errors are small compared to systematic ones estimated \mbox{by~\citet{rackers23};} hence, the latter dominate when combined in quadrature. Thus, the recommended value of $\mu_0$, based on the Best24 sample, is $\mu_0=31.06^{+0.04}_{-0.06}$~mag, and the distance estimate to M87 is $D=16.29^{+0.30}_{-0.45}$~Mpc.

\citet{degrijs20} also discarded one $\mu_0$ value from the Cepheid sample, i.e., the measurement of~\citet{tammann2000}. They argued that this is an outlier; indeed, it has a clearly larger value than others from this tracer and a larger uncertainty as well (see Figure~\ref{fig1}a).~\citet{rackers23} followed suit. I repeated the whole procedure from Section~\ref{sect:2.2} for the two samples from Section~\ref{sect:2.1} without the~\citet{tammann2000} point as well. The only difference, compared to the results quoted in Table~\ref{tab1}, is that the Best40 sample (i.e., the Best41 sample minus this one point) yielded $\braket{m}=30.99$~mag, leading to $D=15.78$~Mpc. Therefore, the~\citet{tammann2000} measurement does not affect the final distance to M87 obtained above, specifically in the case of the Best24 sample.

\subsection{Relevance}
\label{sect:3.2}

The supermassive black hole (SMBH) mass $M_{\bullet}$ scales with distance $D$ as $M_{\bullet}\propto D^2$ \citep{degrijs20}; hence, two mass values $M_{\bullet,1}$ and $M_{\bullet,2}$ corresponding to $D_1$ and $D_2$, respectively, are related to each other as
\begin{equation}
\frac{M_{\bullet,1}}{M_{\bullet,2}} = \left(\frac{D_1}{D_2}\right)^2.
\label{eq3}
\end{equation}
Taking $M_{\bullet,2}=6.5\cdot 10^9 M_{\odot}$ as reported by the~\citet{EHT19} and the $D_2=16.8$~Mpc they employed, the mass $M_{\bullet,1}$ corresponding to the new recommended distance $D_1=16.29$~Mpc is calculated to be $M_{\bullet,1}=6.1\cdot 10^9 M_{\odot}$. The uncertainty $\Delta M_{\bullet,1}$ is not reported here since error propagation\footnote{Note that its application to the mass estimate by~\citet{degrijs20} gives an error of $1.1\cdot 10^9M_{\odot}$, not the $0.6\cdot 10^9M_{\odot}$ reported by them, which was likely obtained by neglecting the terms $\Delta D_1$ and $\Delta D_2$ in the resulting formula for $\Delta M_{\bullet,1}$. Their higher value than that adopted by the~\citet{EHT19} distance uncertainty would, hence, lead to a smaller SMBH mass uncertainty, which is a counterintuitive outcome. Likewise, a similar case occurs 
for the mass estimate obtained with the even smaller uncertainties of the new recommended distance computed herein.} applied to Equation~(\ref{eq3}) would also involve $\Delta D_2$ and $\Delta M_{\bullet,2}$ besides $\Delta D_1$. The mass estimate $M_{\bullet,2}$ by the~\citet{EHT19} was obtained via numerical simulations within a general relativistic framework utilizing just the value and uncertainty of $D_2$. To obtain a meaningful error $\Delta M_{\bullet,1}$ of the SMBH mass estimate, the same simulations would need to be performed independently utilizing the new recommended distance $D_1$ without referring to $D_2$ and $M_{\bullet,2}$.

Finally, the new recommended distance to M87 has significantly smaller errors than previously reported in the literature \citep{rackers23,ramakrishnan23}. This allows us to set M87 more firmly on the cosmic distance ladder as a reference distance, and it aids in setting the next rungs, extending to the $\sim$100~Mpc scale \citep{degrijs20}.

\section{Summary}
\label{sect:4}

Obtaining a representative value for a set of data can be achieved with a measure of central tendency.~\citet{degrijs20} calculated the distance to M87 as the arithmetic mean of absorption-corrected distance moduli $\mu_0$ gathered from the literature published in the last 100 years.~\citet{rackers23} and~\citet{ramakrishnan23} employed the median as the measure of central tendency, obtaining values consistent with each other and with the mean within errors. Herein, I explore the mode of a unimodal distribution as another measure of central tendency. All measures (i.e., mean, median, and mode) are in agreement within errors, but the statistical uncertainty of the mode is greatly reduced compared to those of the medians via bootstrap resampling, and the final error of the mode is dominated by systematic uncertainties established by~\citet{rackers23}. The final recommended distance to M87 is, therefore, $\mu_0=31.06^{+0.04}_{-0.06}$~mag, equivalent to $D=16.29^{+0.30}_{-0.45}$~Mpc.

\vspace{6pt} 

\funding{Funding from the National Science Center through Sonata Grant No. 2021/43/D/ST9/01153 is acknowledged.}

\dataavailability{The original data used in this study are available in Table~1 of \citep{degrijs20}.} 

\conflictsofinterest{The author declares no conflicts of interest.} 

\abbreviations{Abbreviations}{
The following abbreviations are used in this manuscript:\\

\vspace{-6pt}
\noindent 
\begin{tabular}{@{}ll}
KDE & kernel density estimate\\
PDF & probability density function\\
PNLF & planetary nebulae luminosity function\\
SBF & surface brightness fluctuation\\
SMBH & supermassive black hole\\
TRGB & tip of the red giant branch
\end{tabular}
}

\begin{adjustwidth}{-\extralength}{0cm}
\printendnotes[custom]
\reftitle{References}

\PublishersNote{}

\end{adjustwidth}

\end{document}